\documentclass[a4paper,twoside]{article}

\usepackage{epsfig}
\usepackage{subcaption}
\usepackage{calc}
\usepackage{amssymb}
\usepackage{amstext}
\usepackage{amsmath}
\usepackage{amsthm}
\usepackage{multicol}
\usepackage{pslatex}
\usepackage{apalike}
\usepackage[hyphens]{url}
\usepackage{hyperref}
\usepackage{xcolor}
\usepackage{algorithm2e}
\usepackage[bottom]{footmisc}
\usepackage{enumitem}
\usepackage{booktabs}
\usepackage[most]{tcolorbox}
\usepackage{listings}
\usepackage{SCITEPRESS}     % Please add other packages that you may need BEFORE the SCITEPRESS.sty package.
\usepackage{orcidlink}

\newcommand{\cur}{\texttt{.cursorrules}}

\newtcolorbox{myrulebox}{
    colback=gray!5!white,
    colframe=black!75,
    arc=0.5mm,
    boxrule=0.5pt,
    left=2mm,
    right=2mm,
    top=1.5mm,
    bottom=1.5mm,
    boxsep=0pt,
    fontupper=\fontsize{8pt}{9.6pt}\selectfont\ttfamily, 
    sharp corners=northwest
}

\begin{document}

\title{A Study of Cursorrules Files in GitHub Open Source Projects}

\author{\authorname{Shuang Sun\sup{1}, %\orcidAuthor{0009-0009-7585-7213}, 
Jafar Akhoundali\sup{1}, %\orcidAuthor{0009-0002-8260-9508}, 
Arina Kudriavtseva\sup{1}, %\orcidAuthor{0000-0001-7485-0559}, 
Sengim Karayal\c cin\sup{1} %\orcidAuthor{0009-0000-1598-8400} 
and Olga Gadyatskaya\sup{1}%\orcidAuthor{0000-0002-3760-9165}
}
\affiliation{\sup{1}Leiden Institute of Advanced Computer Science, Leiden University, Einsteinweg 55, Leiden, The Netherlands}
%\affiliation{\sup{2}Department of Computing, Main University, MySecondTown, MyCountry}
\email{\{s.sun, j.akhoundali, a.kudriavtseva, s.karayalcin, o.gadyatskaya\}@liacs.leidenuniv.nl}}

\keywords{
%The paper must have at least one keyword. The text must be set in 9-point font and without bold or italics. For more than one keyword, please use a comma as a separator. Keywords must be titlecased.
Cursor, Cursorrules, Prompt Engineering, GitHub, Mining Software Repositories}

\abstract{%The abstract should summarize the contents of the paper and should contain at least 70 and at most 200 words. The text must be set to 9-point font size.
Prompts are the primary mechanism for communicating with AI agents, and they directly influence the quality and reliability of AI-generated code. As AI-assisted programming becomes widely adopted, modern tools increasingly combine dynamic conversational prompts with static configuration-like prompt files. Despite the growing focus on prompt engineering, prior research has primarily focused on conversational prompts, while prompt files remain understudied.
To address this gap, we conduct an empirical study of configuration prompt files in Cursor, a widely used AI-assisted code editor. We collect and analyze over 12,110 \texttt{.cursorrules} files from 11,427 GitHub repositories to characterize their distribution, evolution, and maintenance. Complementing this, we perform qualitative analysis on a random sample of 65 prompt files and develop a 65-code codebook capturing how developers express programming intent, project context, engineering practices, and security considerations.
Our results show that \cur\ files emerged rapidly from mid-2024. Their adoption is concentrated in small-scale, low-activity, single-maintainer repositories, suggesting toy projects rather than professional development. The content of prompt files is dominated by guidance on code quality and engineering practices, project structure and configuration, and maintainability, while security-related content appears less frequently. Our analysis shows that there is a continuity of themes and topics between the now-legacy \cur\ files and the current standard \texttt{.mdc} files. 
%(1) prompt files emerged rapidly from mid-2024 and quickly converged into small, structurally consistent configurations, with adoption concentrated in web-centric, small-scale, and low-activity repositories typically maintained by a single developer.
%(2) Most files remain unchanged after creation, and when updates occur, they are generally small and low-ratio, often involving fewer than ten lines and less than 10\% of the file content, while large or full-content replacements have become less common over time. These files are usually introduced as part of larger batch commits, often within 24 hours of repository creation, whereas subsequent updates are tend to be localized and involve few additional files.
%(3) The content of prompt files is dominated by guidance on code quality and engineering practices (35.17\%), project structure and configuration (20.17\%), and maintainability (10.08\%), while security-related content appears infrequently\seng{Why not percentage here?}.
%(4) Across all codes, guidance on languages and technology stacks is the most common, and several newer file formats sometimes include more detailed contextual explanations or paired examples illustrating both correct and incorrect behaviors. We also identify rare (2.49\%) but present security smells, including exposed local paths, hard-coded parameters, private links, and contradictory instructions.
}

\onecolumn \maketitle \normalsize \setcounter{footnote}{0} \vfill

\section{\uppercase{Introduction}}
% no \IEEEPARstart
%AI-assisted programming has significantly lowered the entry threshold of software development. This enables individuals without a technical background to participate in software development, expanding the popularity of technology creation~\cite{vibe_coding_as_reconfiguration}.
%According to a recent report by Microsoft, the adoption of generative AI nearly doubled in under six months.
AI-assisted programming has become ubiquitous. 
In the 2025 Stack Overflow's annual survey~\cite{stackoverflow2025survey}, $85\%$ of surveyed professional developers reported they are using or are planning to use AI code assistants. These AI assistants are now an integral part of the software development infrastructure, becoming embedded in development environments and introducing artifacts such as prompts and AI service API calls in project files. 

The unprecedented speed of AI coding agents' adoption calls for empirical research into the practices of AI assistants' usage and their influence on software development processes. Specifically, research on \emph{prompts}, the primary mechanism for communication with AI agents, is rapidly increasing.
Prompts have a direct impact on the quality of AI-generated code~\cite{della_prompt_code_quality,nam2025understandingsupportingdevelopersprompt,khojah2025impact}. It is thus important to study prompts used by practitioners as well as propose new techniques to improve prompts for achieving better performance on the target task, which is often referred to as \emph{prompt engineering}.

%Prompts can influence the quality of the produced code~\cite{della_prompt_code_quality,nam2025understandingsupportingdevelopersprompt}, and the recent literature actively investigates how to design effective prompts. \emph{Prompt engineering} refers to various prompt improvement techniques and methodologies that enhance LLM performance in the relevant areas and tasks~\cite{wang2025advanced,chen2025promptware,prompts_are_programms2,ronanki2025prompt,sahoo2024systematic,schulhoff2025promptreportsystematicsurvey,chen2025unleashing}. For example, it was shown that prompts designed for earlier-generation LLMs may not retain their effectiveness for the more advanced models, but execution feedback, for instance, remains effective~\cite{wang2025advanced}. Gao et al.~\cite{gao2023makes} show that sufficiently diverse demonstrations in the prompt improve performance in code intelligence tasks and provide guidance on how to select these samples. 

To get better output, modern AI agents and AI-assisted programming tools often use a hybrid prompt strategy that combines dynamic \emph{conversational prompts} with static \emph{prompt configuration files}.
Conversational prompts are short-lived instructions inside the ongoing conversation between the user and the AI agent.
They influence the agent's behavior and output within the current interaction.
In contrast, static prompt configuration files are relatively stable instructions that operate continuously at a broader, project-level scope.
They provide project context and shape agent behavior for all interactions within the given project, therefore, potentially achieving a deeper impact. While the engineering practices for conversational prompts have already received significant attention in the literature~\cite{chen2025promptware,ronanki2025prompt,sahoo2024systematic,chen2025unleashing,nam2025understandingsupportingdevelopersprompt,siddiq2024quality,siddiq2024fault,mao2025prompts}, the research on prompt configuration files is still limited. \cite{context_engineering} and \cite{chatlatanagulchai2025agent} examined prompt files in open-source projects and analyzed their contents, focusing on \texttt{AGENTS.md}, a recent prompt configuration file format. However, these studies did not reveal how prompt configuration files have been designed by developers \emph{in the wild}, since the emergence of the AI coding assistants using such files.  

%, affect generated code quality and project development, and can trigger cascading effects across the software supply chain.
%Compared with conversational prompts, they have a larger, deeper, and longer-term impact, making them both powerful and risky.
% They affect the SSC by influencing generated code quality, dependency choices, security posture, maintainability, and integration with other tools. 
% They shape agent behavior by defining how the AI interprets tasks, constraints, and style.
% They affect project development by guiding coding patterns, architecture, documentation, and workflow.
%Prompt files are easy to overlook since they appear as ordinary project files, and their influence on specific outputs is often vague and difficult to trace.

\begin{figure}[t!]
    \centering
    \begin{myrulebox}
% --- Your Content Here ---
You are an expert in TypeScript, Node.js, React, and Tailwind.

\vspace{2mm}
Code Style and Structure

- Write concise, technical TypeScript code with accurate examples.

- Use functional and declarative programming patterns; avoid classes.

\vspace{2mm}
Naming Conventions

- Use PascalCase for directories and files (e.g., PixelViewer.tsx).

- Favor named exports for components.
% --- End Content ---
    \end{myrulebox}
    \caption{Example of a \cur\ file.}
    \label{fig:cursor_rule}
\end{figure}

To better understand how such prompt configuration files have been used in practice, we examine the prompt-file ecosystem of Cursor, a widely used code editor built for AI-assisted programming~\cite{cursorDocs}. In our work, we conduct both qualitative and quantitative analyses of the free-form \cur\ files to investigate how prompt configuration files have been adopted by the practitioners, what topics they contain, in which open source projects they appear, and how they have evolved.  %, what they expect from it, and how their interaction patterns change across the development lifecycle. 
We organize our investigation around four research questions. % that explore the content as well as the evolution, maintenance, and contextual characteristics of these files. 

\noindent\textbf{RQ1: What are \cur\ files, and in which open source projects they appear?} Here, we examine \cur\ as files and study in what kind of projects they are located.
\begin{itemize}[leftmargin=*,nosep]
    \item \textbf{RQ1.1}: What is the temporal dynamics of \cur\ files creation in projects?% in open-source GitHub projects?
    \item \textbf{RQ1.2}: How large are \cur\ files and how did their size change over time?
    \item \textbf{RQ1.3}: In which projects are \cur\ files located?
\end{itemize}

\noindent\textbf{RQ2: What is the evolution of \cur\ files?}
\begin{itemize}[leftmargin=*,nosep]
    \item \textbf{RQ2.1}: How are \cur\ files updated?
    \item \textbf{RQ2.2}: What are the characteristics and evolution of commit intervals for \cur? %, and what factors influence them?
    \item \textbf{RQ2.3}: How substantial are the modifications made to \cur?
    \item \textbf{RQ2.4}: Are \cur\ files added or modified independently, and how do their creation and last-update times relate to repository lifecycles?
\end{itemize}

\noindent\textbf{RQ3: What are the major topics in \cur?} Knowing these, we can understand what developers are most concerned with when developing with Cursor.

\noindent\textbf{RQ4: Which security aspects are considered in \cur?} We examine how developers consider security issues and identify risky patterns.

\textbf{Contributions.} We collect a dataset of 12,110 \cur\ files from 11,427 open-source GitHub repositories. %We then examined these files and the containing repositories to understand what kind of projects host Cursor development and how \cur\ files are being maintained. 
Their analysis (\textbf{RQ1\&2}) revealed that most \cur\ files belong to Web application development projects; however, most of these projects are small, unpopular, and are maintained by a single contributor, indicating toy projects rather than large open-source systems. We also performed thematic analysis of 65 random Cursor prompt files, identifying the most common themes and topics that developers care about. This analysis (\textbf{RQ3\&4}) shows that the themes and codes across the two considered file types are %relatively 
similar. Developers prioritize instructions about code quality and engineering practices, as well as project structure and configuration guidance, while security concerns do not receive substantial attention. Our empirical study contributes to a better understanding of the current AI coding assistants ecosystems.

\section{\uppercase{Background}}

% --- 2. cusorrules
At the time of writing, Cursor supports several types of prompt configuration files and refers to them as \emph{rule files} or simply \emph{rules}.
These files are included at the start of the large language model (LLM) context, thereby establishing context persistence and providing consistent guidance across many LLM interaction sessions. Cursor works with four types of rules. Markdown-based \emph{project rules} are stored in \texttt{.cursor/rules}. These files are scoped to the codebase using path patterns and can be version-controlled. They contain metadata specifying the rule application. %These project rules can be stored as \texttt{.mdc} or \texttt{.md} files. 
\emph{User} and \emph{Team} rules are specific to the Cursor AI environment and a user group environment on enterprise plans, respectively. Finally, the fourth type is the universal format \texttt{AGENTS.md}~\cite{agentsmd}, which does not have Cursor-specific metadata, but supports interoperability with other AI agent ecosystems, as it also can be used with Claude, Copilot, Gemini, and others. 

However, previously, Cursor did not have so many rule types, and it used just one format: \cur~\cite{cursorRules}. A representative example of \cur\ is shown in Figure~\ref{fig:cursor_rule}. These \cur\ files were made legacy in February 2025, when Cursor announced that these files would eventually be deprecated, and the users are recommended to move to \texttt{.cursor/rules} or \texttt{AGENTS.md}. However, at the time of writing, they are still supported and persist around the ecosystem.  Studying this earlier format and its emergence can help understand the evolution of prompt configuration files: by examining \cur\ content, we can see what developers care about and understand better the prompt file usage patterns.  %\texttt{AGENTS.md} is a recommended replacement for \texttt{.cursorrules} that defines agent behavior instructions.
%It is a general agent behavior specification file in plain markdown format without metadata, which is adopted by various AI-assisted programming agents and tools, such as Claude, Copilot, Gemini, and others~\cite{agentsmd}.

\cur\ have free formatting; no specific metadata or structure is prescribed, and their content is entirely up to the developers. Given that Cursor has been available since 2023 and is widely used in development~\cite{opsera}, we set out to study the legacy \cur\ files to understand how they have appeared on GitHub and what they contain. 

Beyond \texttt{AGENTS.md} and \cur, similar agent rule file patterns appear in other AI coding assistants, such as \texttt{CLAUDE.md} for Claude Code and \texttt{copilot-instructions.md} for Copilot, reflecting a broader ecosystem trend toward project-level prompt files that requires attention from researchers.

\section{\uppercase{Methodology}}

\begin{figure*}[t]
\centering
\includegraphics[width=0.88\linewidth]{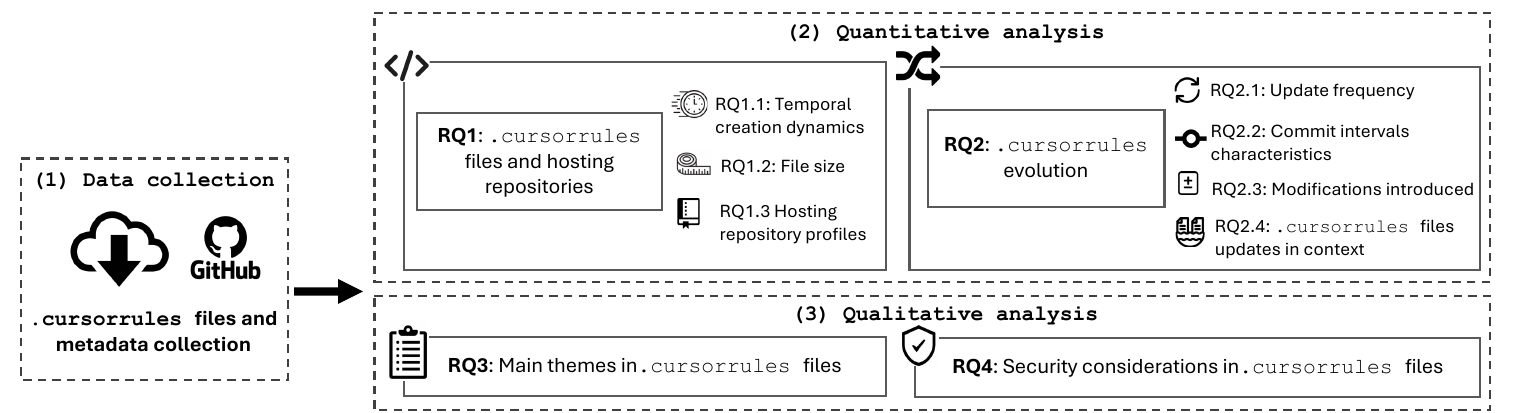}
\caption{Our study methodology. %We first collect all open source software projects with \cur\ files available via the GitHub API, and then we analyze our data quantitatively and qualitatively to answer our four research questions.
} 
\label{fig:methodology}
\end{figure*}

To answer our RQs, we use a mixed-methods approach, which includes both quantitative and qualitative analyses. Our methodology, visualized in Figure~\ref{fig:methodology}, consists of three phases:
(1) data collection, (2) quantitative analysis of the collected data as GitHub code artifacts (answering \textbf{RQ1-RQ2}), and (3) qualitative analysis of the data (answering \textbf{RQ3-RQ4}). 

\subsection{Data Collection}
%\subsubsection{Quantitative Analysis Data Collection}
\textbf{Our dataset.}
%To identify Cursor rule files on GitHub, we used GitHub Code Search with queries targeting \cur\ filenames (e.g., \texttt{filename:.cursorrules}). Because GitHub returns at most 1,000 results per query, we iteratively added additional keywords to partition the search space and ensure that each query produced fewer than 1,000 matches.
%With this strategy, we collected data on two occasions (April and October 2025) and merged the snapshots, keeping the most recent version of each file.
%In total, we obtained 12,110 unique and valid \cur\ files from 11,427 open-source repositories. The dataset covers the period from 15 April 2024 to 30 September 2025.
We used GitHub Code Search to identify Cursor rule files, querying \cur\ filename (e.g., \texttt{filename:.cursorrules}). Because GitHub limits each query to 1,000 results, we iteratively added keywords to partition the search space. We collected data in April and October 2025 and merged the snapshots, keeping the most recent version of each file. In total, we obtained 12,110 unique \cur\ files from 11,427 repositories, covering 15 April 2024 to 30 September 2025.

\textbf{Metadata collection.} 
%To examine the properties of the collected files along with their repositories, we identified 76 data attributes (listed in~\cite{AnonymousGithub}) extracted from GitHub API responses. These attributes characterize a \cur\ file in terms of the file itself, the creator, the commit history, the repository, and the repository owner.
% https://anonymous.4open.science/r/cursorrules/README.md
To examine the properties of the collected files and their repositories, we identified 76 data attributes (listed in our repository~\cite{AnonymousGithub}) extracted from GitHub API responses. These attributes characterize a \cur\ file in terms of the file itself, its creator, commit history, repository, and repository owner.
Some attributes were incomplete due to GitHub's access limits.
For counts such as commits, releases, pull requests, and contributors, we used a threshold of 100 - the maximum returned per API request.
If the true value does not exceed this limit, the crawler records it; otherwise, it returns 101.

For repository attributes like README file, wiki, discussion, and license, we do not request their contents; instead, we check their existence by requesting their fixed-format links and record their presence.
We retrieve commits, pull requests, contributors, and releases via the GitHub GraphQL API, which exposes \texttt{totalCount} fields and reduces the number of API calls.
%For contributor counts, we use the GraphQL \texttt{mentionableUsers.get("totalCount")} query, because the data field \texttt{contributors} from the REST endpoint may be affected by GitHub's privacy policy and return inaccurate values. The \texttt{mentionableUsers} query, on the other hand, counts all users who have access to the repository, providing more reliable results.
%File-level commit changes are significant, as they indicate which lines are added or removed, the content of each modification, and the affected sections.
%To accurately trace how a particular file evolves across commits, we identify all commits that modify the \cur\ file and extract their diffs. Since commit diffs include changes to multiple files, we filter them to isolate only the modifications relevant to the target file.
File-level commit changes show which lines are added or removed and which sections are affected.
To trace a file's evolution, we identify all commits that modify the \cur\ file and extract their diffs. Since diffs may include multiple files, we filter them to isolate only the changes to the target file.

\subsection{Quantitative Analysis}
We first conduct quantitative analyses on the collected dataset. These analyses focus on \textbf{RQ1} and \textbf{RQ2}, and provide a multi-dimensional characterization of \cur\ files and their hosting repositories.

\subsubsection{RQ1: \cur\ files and their open-source projects}
To answer this RQ, we analyze \cur\ file creation times (\textbf{RQ1.1}) and file size dynamics over time (\textbf{RQ1.2}).
To characterize repositories containing \cur\ files (\textbf{RQ1.3}), we examine GitHub-reported language information and repository size, counting the occurrence of each language label across all repositories.
We also analyze repository activity by measuring total commits, monthly commit frequency, and the number of unique contributors. 

%RQ1.3 - (5) etc.
%We further characterize these repositories by analyzing their popularity, social engagement, and completeness.
%Popularity and visibility are measured using the number of stars, forks, and watchers associated with each repository. These metrics are commonly used on GitHub to indicate user interest, adoption, and community attention~\cite{he2024,sun2025beyond}.
%Social engagement was assessed using the number of open issues, which reflects ongoing interactions between users and maintainers.
%For these patterns, we calculate the summary statistics.  
%To evaluate maturity and completeness, we design a composite attribute, \texttt{repo\_functions}, which records the presence of four structural patterns of each repository: a README file, a license, a wiki, and a discussion section.
%We used it as a reference indicator for measuring the completeness of a repository.
%Based on the length of \texttt{repo\_functions}, we grouped repositories and then calculated the percentage of each group.

\subsubsection{RQ2: Evolution of \cur\ files}
%To understand the maintenance activity of \cur\ files (\textbf{RQ2.1}), we first measure whether they are modified and how frequently.
To measure whether \cur\ files are modified and how frequently (\textbf{RQ2.1}), we analyze their maintenance activity by identifying the proportion of files with at least one post-creation commit and examining the distribution of subsequent commits. This quantifies how actively \cur\ files are maintained.

%%%%%%%%%%%%%%
%RQ: commit intervals. modification counts & 30_days creation epochs
%\textbf{RQ2.2: What are the characteristics and evolution of commit intervals for \cur\, and what factors influence them}~\\
For \textbf{RQ2.2}, we define a \emph{commit interval} as the time difference between consecutive commits to the same \cur\ file.
We include only files with at least two commits, extract and sort all commit timestamps, and compute pairwise time gaps. Intervals of zero seconds are excluded.
% (2) 30-day commit interval
To compare maintenance patterns across creation periods while avoiding observation bias, we use a fixed 30-day window after file creation and consider only intervals within this window. We analyze files created between April 2024 and August 31, 2025, ensuring each has a full 30-day observation period.
%Interval distributions were visualized using violin plots with embedded box plots showing medians and interquartile ranges.

%We characterize the nature of \cur\ modifications (\textbf{RQ2.3}) by analyzing both the commit statuses and the magnitude of changes. We first compute the proportion of each commit status (\texttt{added}, \texttt{modified}, \texttt{renamed}, and \texttt{removed}). Next, to quantify the extent of modifications, we analyze the number of lines changed in commits where the status was modified and analyze the scale of modifications by computing the relative change ratio for each commit, defined as the number of lines changed divided by the total number of lines before the commit.
For \textbf{RQ2.3}, we characterize \cur\ modifications by analyzing commit statuses and the magnitude of changes. We compute the proportion of each status (\texttt{added}, \texttt{modified}, \texttt{renamed}, and \texttt{removed}).
We quantify modification extent by examining line changes in modified commits and computing a relative change ratio, defined as lines changed divided by total lines before the commit.

%To determine whether \cur\ additions and modifications occur alone or together with other files (\textbf{RQ2.4}), we examine commits that added or modified a \cur\ file and record the total number of files changed in the same commit.
%To measure when developers add \cur\ to the repository, we compute the difference between each file creation timestamp and the corresponding repository creation timestamp.
%Finally, to assess whether repositories remain active after the last update of their \cur\, we compare the last commit timestamp of each file to that of its repository.
%Files updated within thirty days of data collection were excluded to avoid right-censoring bias caused by ongoing projects.
For \textbf{RQ2.4}, we examine whether \cur\ additions and modifications occur alone or with other files by counting co-changed files in each commit.
We measure when developers add \cur\ by comparing each file's creation timestamp to that of its repository, and assess post-update repository activity by comparing the last commit timestamps of the file and repository.
Files updated within thirty days of data collection were excluded to avoid right-censoring bias.

%%%%%%%%%%%%%%%%%%%%%%%%%%%%%%%%%%%%%%%%%%%%%%%%
%%%%%%%%%%%%%%%%%%%%%%%%%%%%%%%%%%%%%%%%%%%%%%%%
%%%%%%%%%%%%%%%%%%%%%%%%%%%%%%%%%%%%%%%%%%%%%%%%
%%%%%%%%%%%%%%%%%%%%%%%%%%%%%%%%%%%%%%%%%%%%%%%%

\subsection{Qualitative Analysis}\label{sec:qualitative_analysis}

%\subsubsection{Sample}
\subsubsection{Data Analysis}
As manually analyzing all files was infeasible, we randomly selected 65 prompt files -- 51 \cur\ and 14 newer \texttt{.mdc} files. Each file serves as a self-contained artifact showing how developers guide Cursor's AI-assisted programming behaviors.
We performed thematic analysis of the sample following \cite{braun2006using}, identifying recurring patterns of meaning and abstracting them into themes.
This method suits our study well as it allows themes to emerge directly from developers' expressions without predefined categories. We followed the approach used in similar software engineering studies~\cite{ebert2022communication,bernardo2024machine}.

\emph{Codebook construction.}
We conducted an open and descriptive coding, using inductive coding strategies~\cite{saldana2021coding}.
A single coder first coded 41 files, producing 136 distinct codes for semantic and functional concepts in the text. The coder then refined them for clarity and boundaries, yielding 101 codes.
Coding was performed manually in Atlas.ti\footnote{\url{https://atlasti.com/}}, without using any AI-assisted features.
Using this initial codebook, another author familiarized themselves with the data and independently coded the same five \cur\ samples, yielding a Krippendorff's Alpha of 0.674.
After discussing disagreements and refining code definitions, we merged overlaps, removed redundancies, and added three codes. The codebook was updated to 65 codes.
A second round of coding on five additional files increased Alpha to 0.897, confirming high consistency~\cite{lombard2002content}. All authors agreed that the codebook had reached conceptual stability.
%\emph{Final coding of \cur.}
With the refined codebook, the first coder re-coded 41 files and coded 10 new files. No new codes emerged, confirming theoretical saturation~\cite{aldiabat2018data} for \cur\ files.

\emph{Cross-type coding.}
%To validate the codebook across the prompt-file ecosystem of Cursor, we further coded 14 \texttt{.mdc} files.
%Unlike \cur\ , \texttt{.mdc} files are equipped with a frontmatter header specifying their description, globs, and other scope-related metadata.
%This naturally gave rise to a new code, \textbf{Rule File Description, Scopes \& Nested Structures}.
%We also observed that some \texttt{.mdc} files contain examples illustrating how to generate prompts or rules, which led to another new code, \textbf{Prompt/Rule Example/Template}.
%Apart from these two additions, no further codes or code modifications emerged, suggesting that the codebook remained theoretically saturated across rule-file types.
To validate the codebook across Cursor's prompt-file ecosystem, we coded 14 \texttt{.mdc} files, whose frontmatter metadata (e.g., description, globs, scopes) led to a new code, \textbf{Rule File Description, Scopes \& Nested Structures}.
Some \texttt{.mdc} files contain examples for generating prompts or rules, giving rise to another new code, \textbf{Prompt/Rule Example/Template}.
No further changes emerged, suggesting the codebook remained theoretically saturated across file types.

\emph{Building the themes.}
%Within thematic analysis, we aimed to highlight recurring patterns and main considerations in how developers express programming or engineering considerations in their instructions to Cursor. To this end, we reviewed and iteratively grouped the codes into themes by identifying patterns and conceptual similarities across files, merging overlapping codes, and splitting codes that encompassed multiple ideas. Each candidate theme was refined through discussion among all authors to ensure coherence and adequate representation in the dataset.
We reviewed and grouped the codes into themes and iteratively refined them through author discussions to ensure coherence and adequate representation. Final themes and representative codes appear in Table~\ref{tab:codebook}.

\section{\uppercase{Results}}
\label{sec:results}
%In this section, we present the results of our empirical study on \cur\ files in GitHub repositories. 
In this section, we present the results of our empirical study on rule files in GitHub repositories.
%Because
As our analyses rely on different metadata fields, the sample sizes vary slightly across analyses. For each analysis, we report the corresponding sample size.

\subsection{Quantitative Analysis Results}\label{sec:quant_results}

\subsubsection{RQ1: What are \cur\ files and in which open source projects they appear}
\begin{figure}[t]
  \centering
  \includegraphics[width=0.48\textwidth]{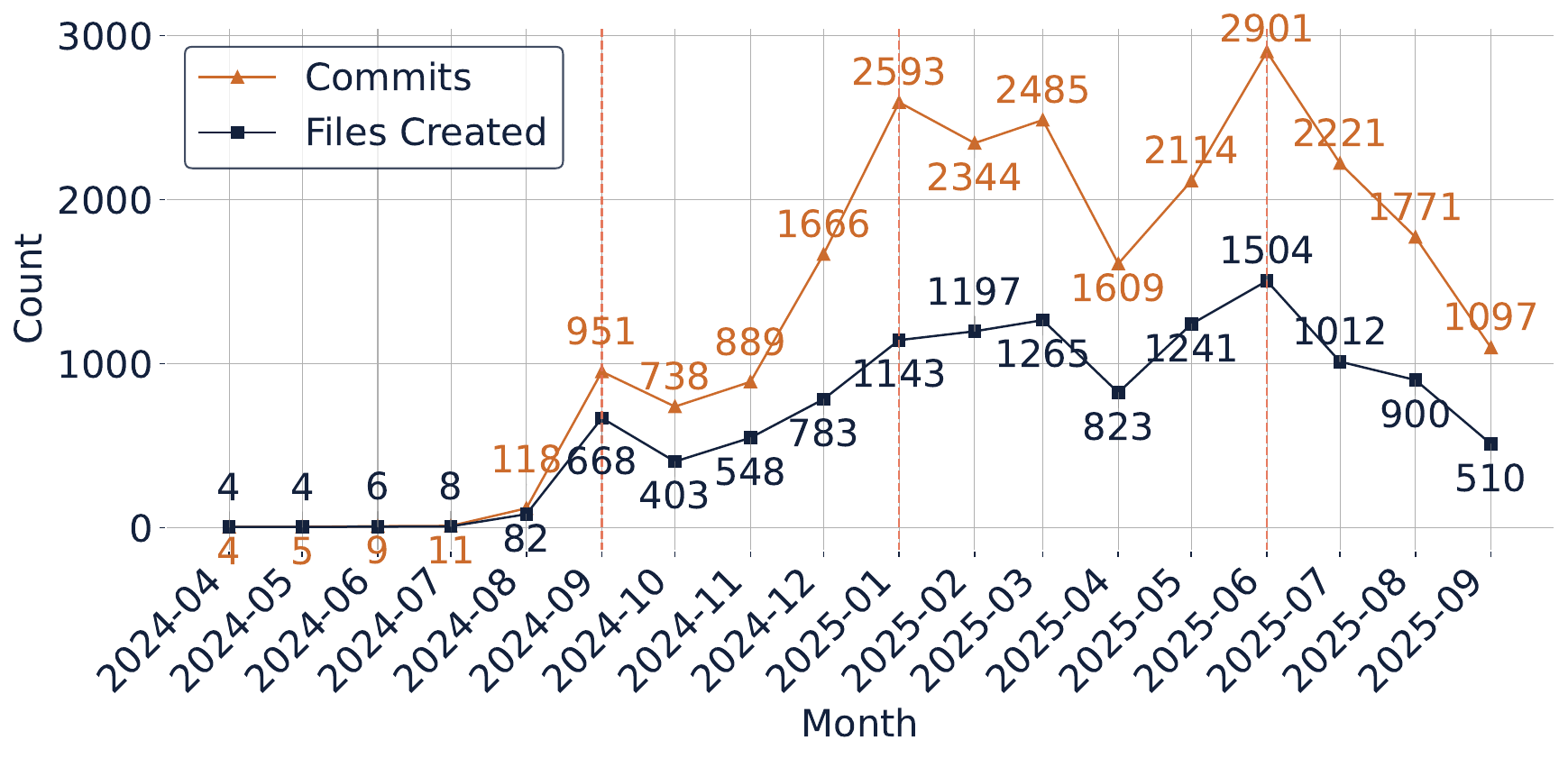}
  \caption{Monthly counts of commits and \cur\ file creations in open-source GitHub repositories.}
  \label{fig:file_creation_and_commit_counts_per_month}
\end{figure}

\textbf{RQ1.1: \cur\ files creation.} 
Figure~\ref{fig:file_creation_and_commit_counts_per_month} shows the monthly distribution of file creations, together with the number of commits for comparison.
The first \cur\ file in our dataset was created on May 29, 2023, and has remained unchanged since then (this outlier is not shown in the figure).
The second file appeared nearly a year later, on April 15, 2024.
From Figure~\ref{fig:file_creation_and_commit_counts_per_month}, we see that \cur\ file creation begins in mid-2024, accelerates toward September 2024, and then stabilizes, with a noticeable decline in the summer of 2025. This decline can be explained by Cursor's transition to a more fine-grained \texttt{.cursor/rules} files, with a general \cur\ file being replaced by \texttt{AGENTS.md}.\footnote{\url{https://cursor.com/docs/context/rules}}

\begin{figure}[t]
  \centering
  \includegraphics[width=0.95\linewidth]{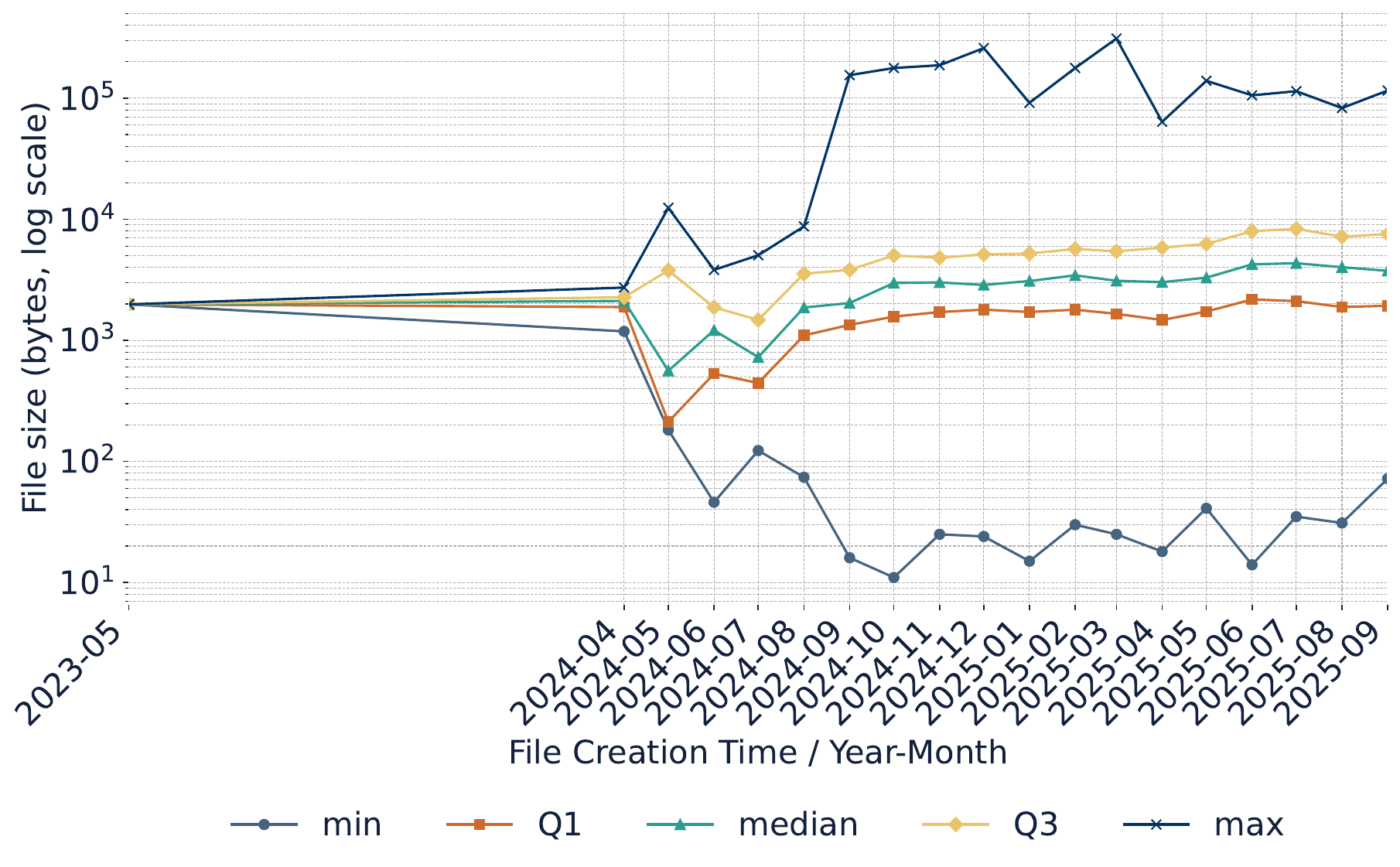}
  \caption{Monthly summary of \cur\ file sizes.}
  \label{fig:file_size_plot}
\end{figure}

\textbf{RQ1.2: Size of \cur\ files.}
%Across 12,085 \cur\ files in our dataset
Across 12,085 sampled files in our dataset, the file sizes range from 11 bytes to 303.67 KB.
The first quartile is 1.73 KB, the median is 3.26 KB, while the third quartile is 5.97 KB, indicating that most files are relatively small while a few are extremely large.
The statistical data for each month are %presented in Table~\ref{tab:file_size_table} (in Appendix~\ref{sec:appendix}) and 
visualized in Figure~\ref{fig:file_size_plot}.
As shown in the figure, all five statistics become more stable from September 2024 onward, while they fluctuated considerably before that period, possibly indicating a gradual stabilization of file sizes over time.
These trends suggest a narrowing range of file sizes and, possibly, more consistent file structures over time.

\begin{figure}[t]
  \centering
  \includegraphics[width=0.95\linewidth]{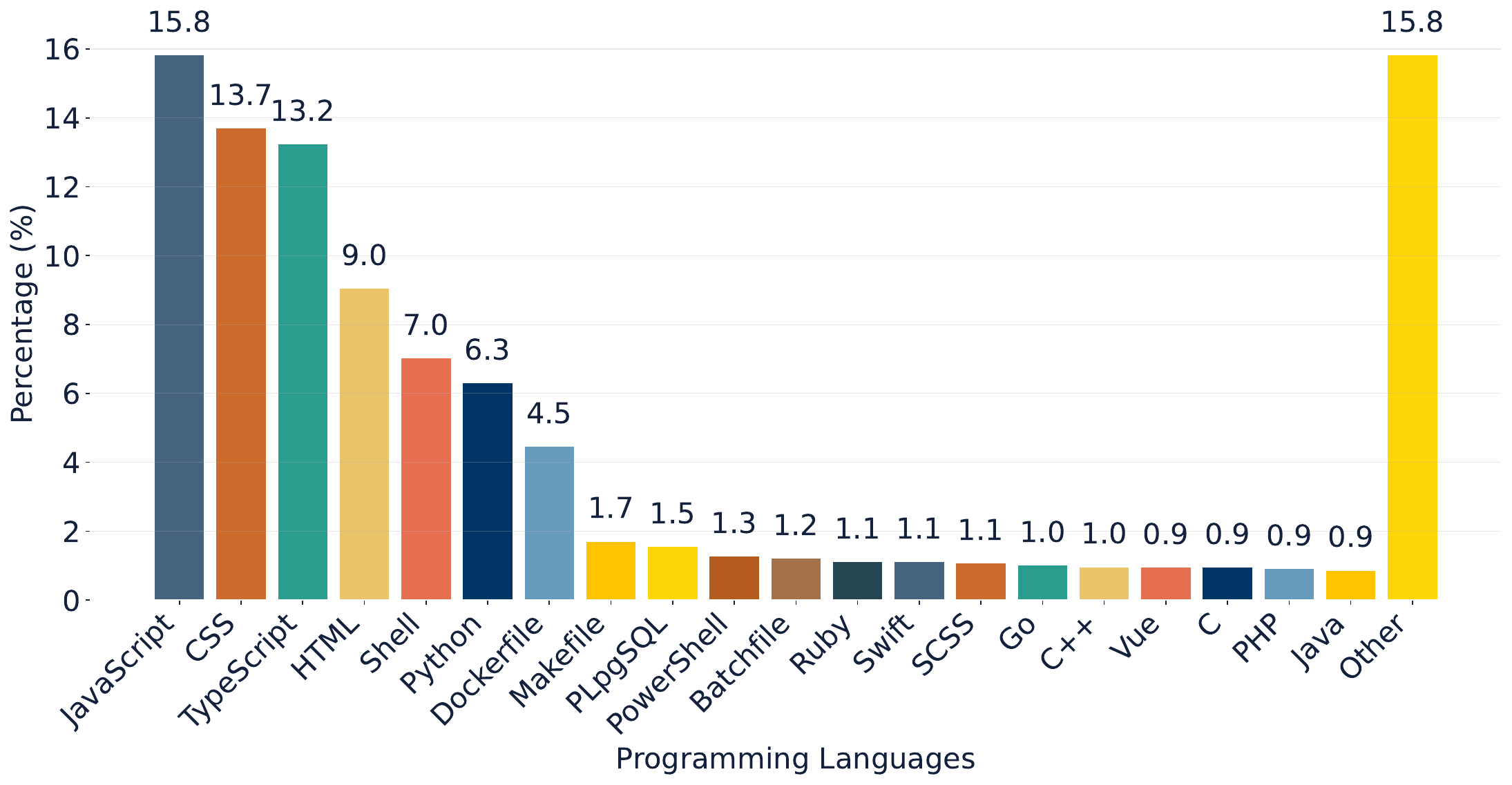}
  \caption{Programming language distribution across repositories containing \cur\ files (in \%). The ``Others'' category aggregates all languages outside the top 20.}
  \label{fig:language_distribution}
\end{figure}

\textbf{RQ1.3: Projects with \cur\ files.}
%We identified 250 distinct programming languages in the repositories containing \cur\ files.
%The top 20 languages account for 84.2\% of all language occurrences, indicating a clear concentration, while the remaining 230 languages constitute a long tail, making up 15.8\% of the total.
%As shown in Figure~\ref{fig:language_distribution}, web development technologies account for more than half, with JavaScript (15.8\%), CSS (13.7\%), TypeScript (13.2\%), and HTML (9.0\%) collectively representing 51.7\% of language usage.
%Shell scripting and Python are also common.
We identified 250 programming languages in repositories containing \cur\ files.
The top 20 languages account for 84.2\% of occurrences, while the remaining 230 form a long tail of 15.8\%.
As shown in Figure~\ref{fig:language_distribution}, web development technologies account for more than half of the usage, with JavaScript (15.8\%), CSS (13.7\%), TypeScript (13.2\%), and HTML (9.0\%) collectively representing 51.7\%.
Overall, \cur\ files appear in a wide variety of project types.

%RQ1.3 - (2) How big are the projects - repo size
%Repository sizes also vary substantially.
%Most repositories are relatively small, with 75.0\% of them under 11 MB, while the maximum reaches 8.76 GB.
%The mean size is approximately 33 MB (SD 177MB), whereas the median is only around 1.36 MB, reflecting a strongly right-skewed distribution driven by a small number of huge repositories.
%The number of files per repository ranges from 1 to 69880. The distribution of file counts across repositories is also highly right-skewed, as can be seen from the difference between the mean of 569 (SD 2854) and the median of 102. Q1 and Q3 are 44 and 252, respectively, resulting in an interquartile range of 208.
%This result shows that \cur\ files are used across projects of various scales, but predominantly in relatively small repositories. We remark that these results might be affected by the fact that we only analyze open-source, public repositories.
Repository sizes and scales also vary substantially.
Most repositories are relatively small, with 75.0\% of them under 11 MB, while the maximum reaches 8.76 GB.
The number of files per repository ranges from 1 to 69880.
The distribution of both sizes and scales has a strongly right-skewed distribution driven by a small number of huge, large-scale repositories.
This result shows that \cur\ files are used across projects of various sizes and scales, but predominantly in relatively small repositories. We remark that these results might be affected by the fact that we only analyze open-source, public repositories.

%RQ1.3 - (3) how frequently they are updated
%The distribution of commit counts among the 11,427 repositories shows that the median commit count is 29, whereas the mean of 213.72 is inflated by a small number of highly active repositories.
%Activity tier distribution shows that 62.5\% of repositories contain 50 or fewer commits, representing low-activity projects. Moderate activity repositories with 51-500 commits comprise 30.3\%, while highly active repositories with more than 500 commits comprise only 7.2\% of the repositories.
%Overall, most repositories experience relatively low development activity, while a minority of repositories generate high commit volumes.
The distribution of commit counts shows a median of 29 commits, whereas the mean of 213.7 is inflated by a small number of highly active repositories.
Activity tiers show 62.5\% repositories contain 50 or fewer commits, 30.3\% have 51-500, and only 7.2\% exceed 500.
%Overall, most repositories exhibit relatively low development activity, whereas a minority exhibit high commit volumes.
%Additionally, for repositories with sufficient temporal data, lifespan $\geq$3 months, we compute their monthly commit frequency to assess the sustained level of activity.
%For the subset of 4,131 repositories with at least three months of lifespan, the median commit rate is 10.34 per month, and the mean is 35.24.
%This confirms the persistence of skewness in commit activity even among repositories with sufficient temporal coverage.
%RQ1.3 - (4) how many maintainers
%Repositories were categorized into tiers based on contributor count: 1, 2, 3, 4-5, 6-10, 11-20, 21-50, 51-100, and $\geq$101.
%To understand the distribution of single-maintainer and team-maintained projects, we further calculate the proportion of single-maintainer repositories, team-maintained repositories, and large teams with more than ten contributors. 
%Among the 11,427 repositories, 68.5\% are maintained by a single contributor, while 31.5\% involve multiple maintainers.
%Projects with two or three contributors account for 17.2\% and 5.0\%, respectively, whereas those with more than five contributors represent less than 8.0\% in total.
%Only 3.4\% of repositories involve large teams with more than ten contributors.
%These results indicate that most repositories remain small in scale and are typically managed by one or a few individuals.
To analyze maintainer structures, we compute proportions of single-maintainer, team-maintained, and large-team repositories. 
Of 11,427 repositories, 68.5\% are maintained by a single contributor, while 31.5\% involve multiple maintainers.
%Projects with two or three contributors account for 17.2\% and 5.0\%, respectively, whereas those with more than five contributors represent less than 8.0\% in total.
Only 3.4\% repositories involve large teams with more than ten contributors.
Overall, most repositories are small-scale and managed by one or a few individuals.

%RQ1.3 - (5) etc.
%The number of stars, forks, and watchers all had a median and lower quartile of zero, indicating that at least 75.0\% of the repositories had not received any external attention.
%The third quartile values of one for stars and watchers suggest that only a small fraction of repositories achieved minimal recognition.
%However, the maximum value is up to 34,166 stars, which indicates that a few repositories are particularly popular.
%A similar pattern was observed in the number of open issues.
%Most projects had little or no active issue tracking, and only a few projects exhibited substantial community activity, with a maximum of 1595 open issues.

%In terms of maturity and completeness, 57.2\% of repositories contained two structural components, of which 92.2\% are a README and a wiki.
%Approximately 22.9\% included three components, while 17.3\% contained only one.
%Cases with all four components and those with none are both rare, accounting for 1.4\% and 1.3\% respectively.

%\highlight{
\underline{\textbf{RQ1 Takeaway:}}
%\cur\ files start to appear at scale from mid-2024. They appear primarily in web-oriented repositories (51.7\%), which are typically small, low-activity, single-maintainer projects with little community attention, despite a long tail of large and popular repositories.
\cur\ files scale up from mid-2024, appearing primarily in web-oriented repositories (51.7\%), which are typically small, low-activity, single-maintainer projects, despite a long tail of large and popular repositories.

\subsubsection{RQ2: What is the evolution of \cur\ files?}
\textbf{RQ2.1: \cur\ file updates.}
%RQ2.1 - (1) How many cursorrules were ever modified?
%\textbf{File modification after creation and monthly commit activity}~\\
%\indent \emph{File modification after creation.}
Among all collected \cur\ files, 67.3\% were never changed after creation, while 32.7\% were modified at least once.
%Table~\ref{tab:mod_count_all} shows the distribution of modification counts.
Most modified files were changed only once (16.1\%), while a smaller proportion (5.1\%) had more than four updates.

%\begin{table}[t]
%\centering
%\caption{\cur\ files modifications.}
%\label{tab:mod_count_all}
%\scalebox{.8}{
%\begin{tabular}{lcc}
%\toprule
%\# of Modifications & Count & Percentage \\
%\midrule
%0 & 8,147 & 67.3\% \\
%1 & 1,953 & 16.1\% \\
%2 & 745 & 6.2\% \\
%3 & 414 & 3.4\% \\
%4 & 228 & 1.9\% \\
%$>$4 & 616 & 5.1\% \\
%\bottomrule
%\end{tabular}
%}
%\end{table}

%RQ2.1 - (2) How often are they updated?
%\emph{Monthly commit activity.}
As shown in Figure~\ref{fig:file_creation_and_commit_counts_per_month}, the number of commits involving \cur\ files increased markedly over time.
From only a few commits per month in mid-2024, activity rose steadily, surpassing 2,000 commits per month in January 2025 and peaking at over 2,900 in June 2025.
Thus, the usage of \cur\ files is rising steadily over time.

%RQ2.1 - (3) *How many people are involved in updating and maintaining cursorrules files?
%\emph{Number of maintainers per file.}
The number of distinct contributors per file exhibits a strong skew. 
Most \cur\ files (95.8\%) were handled by a single contributor. 
Files with two contributors account for 3.9\%, while those with three or more contributors represent only 0.3\% combined. 
Compared with the relatively lower incidence of single-maintainer repositories in our dataset, this indicates that \cur\ files are typically managed individually.

\textbf{RQ2.2: Commit intervals}
%\indent \emph{Commit interval vs. modification counts.}
Table~\ref{tab:commit_intervals_statistics} summarizes commit interval statistics grouped by modification count.
Files edited only once show a median interval of 2.20 days, while those modified more than four times have a median of 0.74 days.
The mean interval follows a similar trend, decreasing from 22.45 days for single-modification files to 5.23 days for the most frequently updated group.
This shows that files receiving more modifications are updated more regularly and with shorter delays between commits.

\begin{table}[t]
  \centering
  %\caption{Commit interval statistics grouped by modification count (seconds and days).}
  \caption{Commit-interval statistics by modification count.}
  \label{tab:commit_intervals_statistics}
  \scalebox{.68}{
  \begin{tabular}{lcccc}
    \toprule
    Group & Min (s) & Median (d) & Avg (d) & Max (d) \\
    \midrule
    1     & 4 & 2.20 & 22.45 & 315.81 \\
    2     & 6 & 1.22 & 13.35 & 359.01 \\
    3     & 4 & 0.98 & 8.88 & 260.10 \\
    4     & 17 & 0.96 & 8.71 & 279.60 \\
    $>$4  & 1 & 0.74 & 5.23 & 223.51 \\
    \bottomrule
  \end{tabular}
  }
\end{table}

%\emph{Commit intervals vs. creation epochs.}
%Figure~\ref{fig:first_30d_commit_interval_violinplot} shows how commit intervals changed between June 2024 and August 2025.
%We further examined how commit intervals changed between June 2024 and August 2025.
%The results show that, between June and December 2024, the median interval dropped from 189.4 hours to 16.0 hours, a 91.5\% decrease. This marks a shift from weekly to near-daily maintenance.
%From January 2025 onward, intervals remained stable, mostly between 13 and 19 hours. This suggests that \cur\ updates became part of regular daily development rather than occasional project-level changes.
We also examined temporal trends. The median interval decreased from 189.4 hours in June 2024 to 16.0 hours in December 2024 and then stabilized between 13 and 19 hours from January 2025 onward.
The overall mean interval (71.0 hours) stayed much higher than the median (18.7 hours), indicating that once a \cur\ file is updated, it is usually updated frequently.

%\begin{figure}[t]
%  \centering
%  \includegraphics[width=0.95\linewidth]{figures/first_30d_commit_intervals_per_file.pdf}
%  \caption{Commit interval distribution within 30 days of \cur\ creation, grouped by file creation month.}
%  \label{fig:first_30d_commit_interval_violinplot}
%\end{figure}

\textbf{RQ2.3: Size of modifications.}
The commit status distribution shows that most updates were new file additions or modifications to existing ones, with few deletions or renames.%with relatively few deletions or renames.
Specifically, 50.2\% of commits are \texttt{added} and 48.0\% \texttt{modified}, while only a small fraction are \texttt{renamed} (1.4\%) or \texttt{removed} (0.5\%).

\begin{figure}[t]
    \centering
    \includegraphics[width=0.38\textwidth]{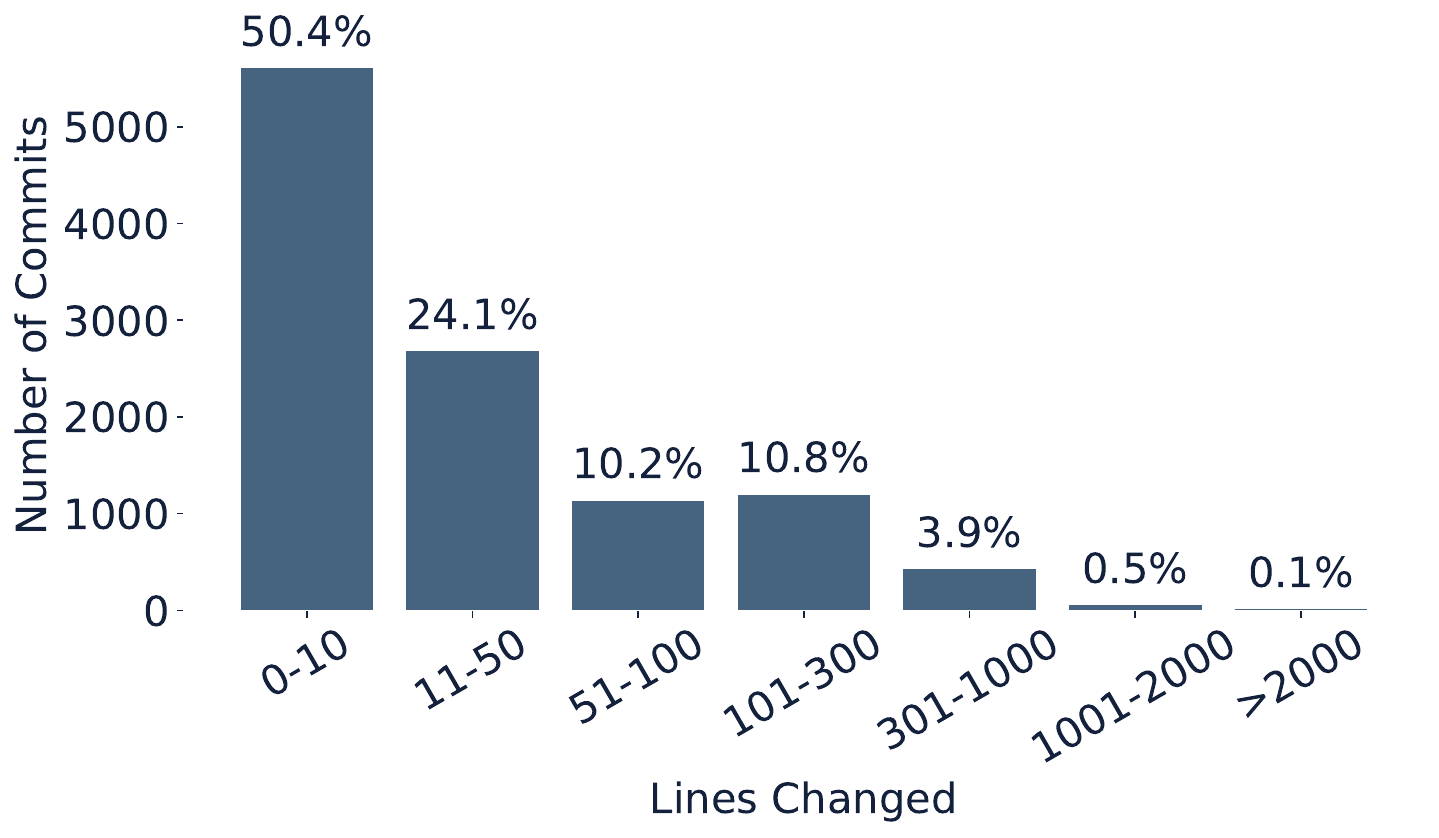}
    \caption{Distribution of the number of lines changed per modified commit. % Most commits involve small-scale edits, with over half affecting fewer than ten lines.
    }
    \label{fig:changed_LoC}
\end{figure}

%\emph{Lines changed per commit.}
Figure~\ref{fig:changed_LoC} illustrates the distribution of the number of lines changed in modified commits.
Over 50.4\% of all commits changed fewer than 10 lines, and 74.5\% changed fewer than 50 lines.
%Only a very small portion of commits (0.6\%) 
Only 0.6\% of commits modified more than 1,000 lines. Therefore, most updated to \cur\ are minor.

\begin{figure}[t]
    \centering
    \includegraphics[width=0.48\textwidth]{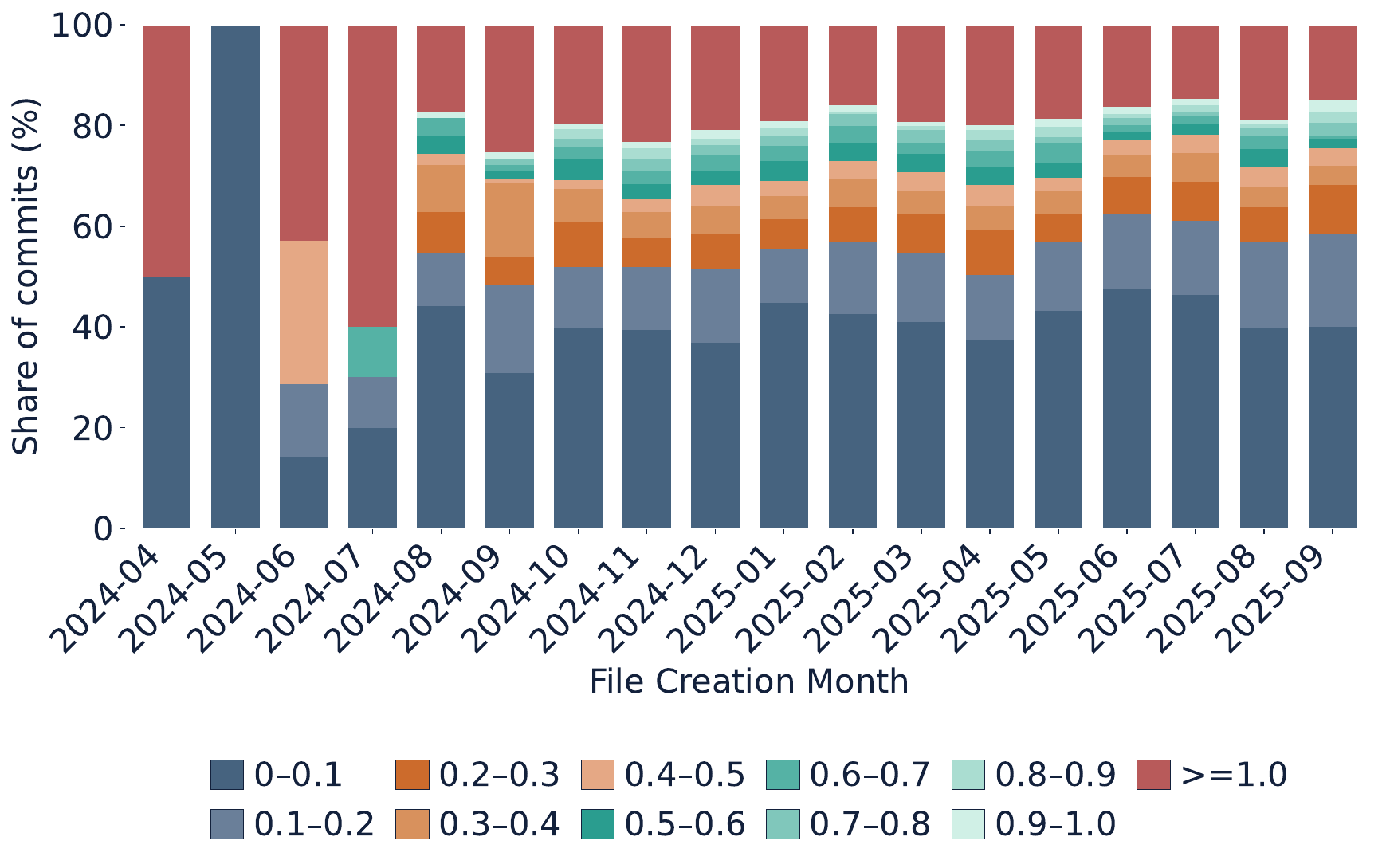}
    \caption{Monthly distribution of \cur\ commits by relative change ratio. Each color interval represents a ratio range. %Most commits involve small relative changes, though large revisions occur periodically.
    }
    \label{fig:change_ratio_bars}
\end{figure}

%\emph{Relative change ratio.}
%We further examined how the number of changed lines compares to the total size of each file.
%As shown in Figure~\ref{fig:change_ratio_bars}, we grouped ratios into 11 intervals, and aggregated the distribution of commits across these intervals by file creation month, covering the period from April 2024 to September 2025.
%The majority of commits involved small proportional changes, typically less than 10\% of the file (ratio$<0.1$).
%The portion of the ratio $>1.0$, which means that the number of changed lines was greater than the total file size before modification, decreased from 2024 to 2025.
%We can observe some trends between 2024 and 2025. 
%First, the yearly averages of the monthly shares for the 0-10\%, 10-20\%, 20-30\%, and 30-40\% bins are all higher in 2025 than in 2024.
%This indicates that commits in 2025 were more likely to involve low-ratio (under 40\%) modifications. 
%Second, high-ratio modifications (ratio $>$50\%) decreased overall in 2025.
%Table~\ref{tab:ratio_combined} summarizes the aggregated statistics of relative-change ratios.
%As shown in the table, 
%The average total share across the six high-ratio bins is 27.46\% in 2025 versus 35.13\% in 2024.
%The decline is most pronounced in the extreme category ($\geq$100\%): its average share fell from 28.87\% in 2024 to 17.57\% in 2025.
%This reduction suggests fewer wholesale replacements of file content, and a trend toward higher stability and more incremental maintenance.
We further examined how the number of changed lines compares to the total size of each file.
As shown in Figure~\ref{fig:change_ratio_bars}, most commits involved small proportional changes, typically under 10\%. Large changes (ratio $>1.0$), where the modified lines exceeded the file's prior size, became less common from 2024 to 2025. Overall, the share of high-ratio modifications declined, especially in the $\geq$100\% category, indicating a shift toward more incremental updates.

\textbf{RQ2.4: Relationship with the repository lifecycle.}
%(1) Co-modified files per commit; (2) File creation time vs. repository creation time; (3) File last update vs. repository last update.
%\indent \emph{Co-modified files per commit.}
As shown in Table~\ref{tab:other_files_when_added}, when \cur\ files were added, 60.7\% of the commits affected more than ten files, only 14.9\% added \cur\ independently.
This suggests that many \cur\ additions occur as part of batch operations.
Similarly, when \cur\ are modified, 19.4\% of commits changed only the \cur\ file, and 60.2\% modified at most 10 files.
%Thus, subsequent edits are more often localized changes rather than bulk operations.

\begin{table}[t]
  \centering
  \caption{Distribution of file counts in commits involving \cur\ files.}
  \label{tab:other_files_when_added}
  \scalebox{.76}{
  \begin{tabular}{lccc}
    \toprule
    Commit Type & 1 & 2-10 & $\textgreater$10 \\
    \midrule
    \texttt{added} & 472 (14.9\%) & 773 (24.4\%) & 1926 (60.7\%) \\
    \texttt{modified} & 566 (19.4\%) & 1194 (40.8\%) & 1163 (39.8\%) \\
    \bottomrule
  \end{tabular}
  }
\end{table}

%\emph{File creation timing relative to repository creation.}
We next analyzed the delay between repository creation and \cur\ creation.
As shown in Table~\ref{tab:init_create}, about 40.7\% of \cur\ files were added within the first 24 hours after repository creation, while 31.7\% were introduced more than 30 days later.

\begin{table}[t]
\centering
\caption{Delay between \cur\ creation and repository initialization.}
\label{tab:init_create}
\scalebox{.76}{
\begin{tabular}{lrr}
\toprule
Time Interval       & \# Files & Percentage \\
\midrule
0{\textless}${\Delta}t{\leq}24$ hours        & 3856      & 40.7\%     \\
24h{\textless}${\Delta}t{\leq}7$ days          & 1414       & 14.9\%      \\
7{\textless}${\Delta}t{\leq}30$ days         & 1210      & 12.8\%     \\
${\Delta}t${\textgreater}30 days         & 3001      & 31.7\%     \\
\bottomrule
\end{tabular}
}
\end{table}

To investigate immediate additions in more detail, we analyzed the first three hours after repository creation in five-minute intervals. %Figure~\ref{fig:creation_difference_less_3_hours} shows that 63.4\% of those files created within three hours are added within the first 20 minutes, and 82.8\% are added within the first two hours.
The result shows that 63.4\% of those files created within three hours are added within the first 20 minutes, and 82.8\% are added within the first two hours.
This indicates that when developers add \cur\ immediately after creating a repository, they typically do so within minutes rather than hours.

%\begin{figure}[t]
%\centering
%\includegraphics[width=0.48\textwidth]{figures/creation_difference_less_3_hours.pdf}
%\caption{Fine-grained distribution and cumulative share of \cur\ creations within the first three hours after repository creation. Bars show file counts per five-minute interval. The line shows the cumulative percentage.}
%\label{fig:creation_difference_less_3_hours}
%\end{figure}

\begin{figure}[t]
\centering
\includegraphics[width=0.48\textwidth]{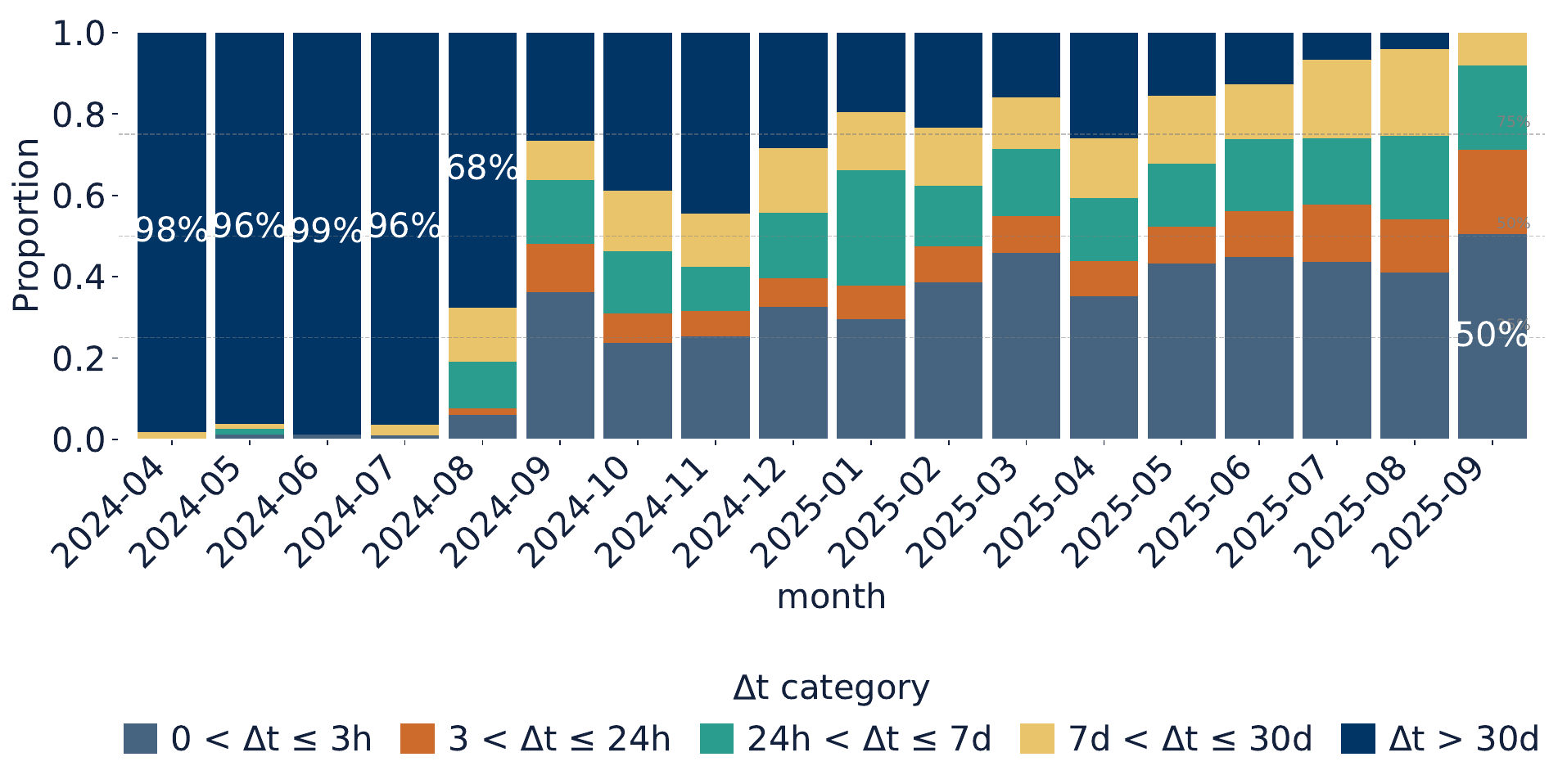}
\caption{Monthly trend of creation-timing categories ($\Delta t$) for \cur\, April 2024 - September 2025. %The share of near-immediate additions increases over time.
}
\label{fig:creation_difference_trend_per_month}
\end{figure}

We also examined monthly trends of $\Delta t$.
Figure~\ref{fig:creation_difference_trend_per_month} reports monthly shares of creation timing categories from April 2024 through September 2025. The share of files created more than 30 days after repository initialization decreased substantially over this period, while the share created shortly after repository initialization increased. This shift suggests growing developer awareness or adoption of \cur\ during repository setup over time.

%\emph{Repository activity after the last \cur\ update.} 
Finally, comparing file last-update times with repository last-update times shows that most repositories remain active after a \cur\ update. Among 11,780 files, 2.6\% of repositories had no updates after the file's last change, indicating that the file was the final recorded modification. A majority of repositories, 54.2\%, received further updates within 30 days of the \cur\ update, and 43.3\% remained active for more than 30 days after the update.
These figures indicate that \cur\ updates commonly occur within active projects and are often part of continued development.

%\highlight{
\underline{\textbf{RQ2 Takeaway:}}
\cur\ files are often left unchanged after creation (67.3\%), and most updates to them are small. %High-ratio modifications became less common in 2025.
They tend to be added in batch commits, while subsequent modifications are more localized and involve fewer other files.
Commit activity has risen steadily since mid-2024.

%%%%%%%%%%%%%%%%%%%%%%%%%%%%%%%%%%%%%%%%%%%%%%%%%%%%%%%%%%%
\subsection{Qualitative Analysis Results}\label{sec:qual_results}

\begin{table*}[t]
\centering
\caption{Themes and Representative Codes}
\label{tab:codebook}
\renewcommand{\arraystretch}{1.1}
\setlength{\tabcolsep}{4pt}
\scalebox{.6}{
\begin{tabular}{p{0.32\linewidth} p{0.36\linewidth}| p{0.32\linewidth} p{0.36\linewidth}}
\toprule
\textbf{Theme} & \textbf{Representative Codes} & \textbf{Theme} & \textbf{Representative Codes} \\
\midrule
\textbf{Access Control} &
Access Control \newline
Authentication \& Authorization \newline
Token Handling \newline
Role-Based Access Control 
&
\textbf{Architecture and Integration} &
Dependency Management \& Injection \newline
HTTP Architecture \newline
Modular Architecture \& Scalability \newline
Routing \\
\hline
\textbf{Agent Behaviors and Rule Files Structure} &
Exclusion Behaviors \newline
Cursor Role Definition \newline
Rule File Description, Scopes \& Nested Structures &
\textbf{Code Quality and Engineering Practices} &
Coding Example/Template \newline
Best Practices \newline
Coding Style \newline
Code Review \\
%Performance optimization \newline \\
%Type safety and consistency \\
\hline
\textbf{Project Structure and Configuration} &
Languages, Stacks \& Usage Guidance \newline
Project Structure \& Architecture \newline
Project Context \& Explanation \newline
Project \& Environment Configuration &
\textbf{Maintainability and Evolution} &
Code Readability \& Maintainability \newline
Comment \newline
Reusability \newline
System Currency \& Iteration \\
\hline
\textbf{Build, Deployment, and CI/CD} &
CI/CD \newline
Deployment \& Release Management &
%Build exclusion rules \newline
%Reference Documentation \& Libraries &
\textbf{Security} &
Security Awareness, Mechanisms \& Practices \newline
Security Smells \\
%Access control mechanisms \newline
%Input sanitization \\
\hline
\textbf{System Input, Data Flow, and System State} &
Database Design \& Operations \newline
Validation \& Sanitize Guidance \newline
State Mangement \& Consistency \newline
Data Fetching &
\textbf{Testing and Error Handling} &
Error \& Exception Handling \newline
Testing \newline
Logging \& Monitoring/Tracking \newline
Debugging \& Troubleshooting \\
\bottomrule
\end{tabular}
}
\end{table*}

\subsubsection{RQ3: What are the major topics in \cur?}
Theme \textbf{Code Quality and Engineering Practices} accounts for the largest share at 30.4\%.
Typical guidance in this theme instructs Cursor about tooling, performance considerations, coding style, naming conventions, and general best practices. For example, ``\textit{Write clean, simple, readable code and follow the project's naming conventions.}''
%%% New Result from .mdc files:
The member code \textbf{Coding Example/Template} captures cases where developers provide examples or templates to help Cursor better understand the programming requirements and generate more appropriate code.
In the sampled \cur\ files, these examples are typically single positive demonstrations.
In contrast, several \texttt{.mdc} files begin to include paired examples, a correct version alongside a wrong one, providing both positive and negative guidance to more explicitly shape Cursor's behavior. For example, ``\textit{Correct Example (GitHub compatible): ..., Incorrect Example (GitHub incompatible due to icons): ...}''.
Themes and representative codes are presented in Table~\ref{tab:codebook}, with the full codebook and top 10 codes in both \cur\ and \texttt{.mdc} files are available in~\cite{AnonymousGithub}. %\olg{could you please put the tables there? there is no space in the paper}\shu{Done.} % Table~\ref{tab:cursor_top_codes} presents the top ten codes in \cur\ files. %Among the themes, the code occurrences are distributed unevenly.
%\shu{Arina: (1)The numbers don't need to be accurate to two decimal places; whole numbers are enough. (5) Consider not using specific percentage values, but only statements such as top 5, top 10.}

\textbf{Project Structure and Configuration} at 18.3\% is the second-largest theme, reflecting that developers commonly specify languages, stacks, and directory layout. For example, ``\textit{This project uses FastAPI for the backend and Python 3.11, organizes modules under the services directory.}''
%%% New Result from .mdc files:
Beyond these general patterns, the member code \textbf{Project Context \& Explanation} shows an additional trend in the sampled \texttt{.mdc} files tend to provide more explanatory information.
For example, ``\textit{This project runs on a vite-based node server ... This means that ... the server will already be running and you don't need to re-run it. This also means that you will avoid at all costs to automatically create new live server ...}''
Additionally, several \texttt{.mdc} files offer specific guidance related to path-related behaviors. %, an area where AI-generated code commonly makes mistakes.
For example, ``\textit{All file paths must be relative or use full URL imports}''. %, ``\textit{Provide Real File Links - Always provide links to the real files ...}''

\textbf{Agent Behavior \& Rule Files Structure} at 14.1\% is the third largest theme, which regulates behaviors of Cursor rather than the code output, e.g., rules for what to do, how to do it, in what role, and within what contextual assumptions. 
Member codes \textbf{Rule File Description, Scopes \& Nested Structures} and \textbf{Prompt/Rule Example/Template} newly emerge in the sampled \texttt{.mdc} files, results in a slightly higher proportion of this theme in \texttt{.mdc} files (18.4\%) compare to \cur\ files(13.0\%).
%Developers frequently remind Cursor to prioritize readability, add helpful comments, and keep code easy to maintain. For example, ``\textit{Favor small functions with clear names, and include comments explaining non-obvious logic.}''

Content from the theme \textbf{Security} is less frequent at 4.2\%. This includes both security design considerations and what we identified as potential security smells.
However, the code \textbf{Security Awareness, Mechanisms \& Practices} shows a clear difference between sampled \cur\ and \texttt{.mdc} files.
In \cur\ files, it is the ninth most frequent code at 3.0\%, whereas in \texttt{.mdc} files it drops to 0.6\% and ranks 41st.
This difference may be partly due to their different scopes: \cur\ files are unrestricted, while each \texttt{.mdc} file has a defined scope. The sampled \texttt{.mdc} files may simply not be security-related, which could contribute to the lower frequency of security content.
%Table~\ref{tab:theme_freq} reports the relative share of coded instances by theme and the average frequency per code within each theme.

Other themes emerged with lower prevalence, but they remain noteworthy. The \textbf{Access Control} theme describes rules that integrate identity, permissions, and resource boundaries directly into the project. Common instructions like ``\emph{Implement proper authentication and authorization}'', ``\emph{Use proper token permissions}'', and ``\emph{Validate user permissions before any operation}'' focus on authentication and authorization mechanisms. Developers also describe how access control should be implemented in different environments. For example, for Laravel, ``\emph{use built-in features like Sanctum, Policies, and authentication scaffolding}''. %; for Next.js, Clerk-based authentication is required (import { auth } from \texttt{@clerk/nextjs/server}).

\textbf{Build, Deployment, and CI/CD} emphasizes following official documentation and best practices for frameworks, libraries, and deployment tools (React, Next.js, Vue.js, FastAPI, Vercel AI SDK, 
PyTorch, Transformers, etc.). 
Instructions enforce standardized workflows, CI/CD (GitHub Actions, ArgoCD), containerization (Docker, Knative), build/test commands, and file handling.

\textbf{System Input, Data Flow, and System State} focuses on safe, efficient, and reliable data and state management. Key points that developers mention include input validation, secure handling of sensitive data, proper database transactions and caching, structured state management, and systematic file and data pipeline handling to ensure integrity and performance. For example, Pydantic should validate inputs, 
%sensitive data should be encrypted, 
transactions should be handled atomically, caching should be used for speed, etc. %, and state should be managed via useState or useReducer; files and pipelines should follow strict validation.

\textbf{Architecture and Integration} describes modular, scalable, and maintainable architectures with clear separation of concerns, and reusable components. Key ideas developers mention include dependency management, API versioning, repository patterns, event/listener decoupling, dynamic routing, and consistent route organization. Integration guidance covers combining frontend and backend components, % (e.g., Livewire + Alpine.js, React Server/Client Components, Nuxt/\$fetch),
managing shared state, middleware, data pipelines, and handling dependencies securely and up to date. 

\textbf{Maintainability and Evolution} focuses on code readability and clear documentation. Code should be readable, self-documenting, and concise, with descriptive names, early returns, and minimal duplication. Projects should be structured into small, focused files with reusable components, logical organization, and separation of configuration, and robust version control practices should be used. %, such as conventional commits and linear history, support traceability and collaboration. Overall, these instructions ensure that code is robust, scalable, and easy to maintain over time.

\textbf{Testing and Error Handling} includes prioritizing early error detection using guard clauses and early returns, implementing proper exception handling with clear and user-friendly messages, creating custom error types when needed, and logging errors for debugging and monitoring. It also includes testing strategies such as unit, integration, and end-to-end tests, proper async testing, test-driven development where appropriate, high coverage on critical paths, and the use of tools like PHPUnit, Laravel Dusk, Jest, Playwright, or go test to ensure reliability and maintainability.

%To give a more concrete sense of content at the code level, the top codes by frequency are summarized in Table~\ref{tab:top_codes}.

In terms of specific codes, \textbf{Languages, Stacks and Usage Guidance} is the most frequent code in both \cur\ and \texttt{.mdc} files, which instruct the coding assistant on the specific languages and stacks to use and which persona to adopt, e.g., instructing it to act as ``\textit{FastAPI backend developer with deep knowledge of Python 3.11''} or \textit{``Expert in TypeScript, Node.js and React.}''
\textbf{Exclusion Behaviors} is also common, shaping Cursor's behavior through negative constraints such as forbidden actions or behaviors to avoid. These rules direct Cursor to generate more compliant output. For example, ``\textit{DO NOT write complicated and confusing code}''.
\textbf{Coding Example/Template} shows over 3 times higher proportion in \texttt{.mdc} (6.8\%) than in \cur\ files (1.9\%).
Given that the sampled \texttt{.mdc} files also introduce paired examples, this difference may suggest that their coding examples are both more developed and more frequent than those in the sampled \cur\ files.
\begin{table}[t]
\centering
\caption{Theme-level distribution of coded instances and average frequency per code.}
\label{tab:theme_freq}
\scalebox{0.6}{
\begin{tabular}{p{0.96\linewidth} p{0.28\linewidth} p{0.24\linewidth}}
\toprule
%Theme & Share of coded instances& Avg. per code \\
Theme & Share (\%) & Avg./code \\
\midrule
Code Quality and Engineering Practices & 30.4\% & 2.51\% \\
Project Structure and Configuration & 18.3\% & 3.36\% \\
Maintainability and Evolution & 14.4\% & 1.44\% \\
Cursor Behaviors and Role Context & 8.2\% & 1.63\% \\
System Input, Data Flow and State Management & 7.3\% & 0.67\% \\
Testing and Error Handling & 6.0\% & 1.00\% \\
Architecture and Integration & 5.8\% & 1.44\% \\
Security & 4.2\% & 1.80\% \\
Build, Deployment, and CI/CD & 2.8\% & 0.69\% \\
Access Control & 1.0\% & 0.14\% \\
\bottomrule
\end{tabular}
}
\end{table}
%}\fi

%\highlight{
\underline{\textbf{RQ3 Takeaway:}}
Code quality and engineering practices dominate the sampled rule files (30.4\%), with Language, Stacks \& Usage Guidance as the most frequent individual code. %Security-related content appears infrequently overall (4.2\%). 
Coding examples differ by format, with \texttt{.mdc} files containing more frequent examples and introducing paired ones.
%}

\subsubsection{RQ4: Which security aspects are considered in \cur?} 
\indent We use the code \textbf{Security Awareness, Mechanisms \& Practices} to capture explicit best practices, experience, and awareness expressed in rule files.
This content appears with multiple security aspects, but it is relatively infrequent, accounting for roughly 2.5\% across all sampled rule files (including \texttt{.mdc}).

Specifically, when security is discussed, developers typically mention:
(1) security mechanisms and defensive techniques, such as ``\emph{Implement proper CSRF protection and security measures}'', ``\emph{Use Laravel's prepared statements to prevent SQL injection}'', %``Implement proper security measures, including ... XSS prevention ...'', 
or ``\emph{Follow the Checks-Effects-Interactions pattern to prevent reentrancy and other vulnerabilities}''.
(2) input validation and sanitization, such as ``\textit{validate all input data}'' or ``\textit{sanitize all incoming data to prevent injection attacks}'';
(3) secure data handling and storage was also mentioned, such as ``\textit{Use Expo's secure store for sensitive data}'';
(4) general awareness or meta-level reminders without concrete action, such as ``\textit{Security best practices are followed}'', or ``\textit{Security concerns}''.
%This pattern shows that developers sometimes provide concrete security guidance, but some mentions remain brief and general.
%The pattern suggests that developers are sometimes aware of security needs and provide actionable guidance. However, many security mentions are only brief and general. % without detailed operational steps.

Apart from this code, the theme \textbf{Access Control} also captures security-related content and accounts for 1.9\% across the sampled rule files.
This theme consists of 7 member codes (e.g., Authentication \& Authorization, Role-Based Access Control), all focusing on the design and implementation of access control mechanisms, i.e., who can access what resources under what conditions.
Together with \textbf{Security Awareness, Mechanisms \& Practices}, the combined proportion reaches roughly 4.4\%, which suggests that developers are sometimes aware of security needs, but overall treat security considerations only occasionally.

In addition to explicit security guidance, we also identified security smells in the files.
These smells are not direct vulnerabilities but patterns that may indicate risky practices or unintended exposure of internal information.
They include exposed local paths, hard-coded parameters that may represent secrets, links to internal or private resources, and inconsistent or contradictory instructions that could lead to insecure code generation.
Such smells are relatively rare, about 1.5\% of all coded instances, but they are present and indicate that instructions in rule files may unintentionally introduce security risks.
Representative security smell examples are:

(1) \textbf{Information Exposure Smell}. Potential disclosure of internal file paths, configuration details, or system metadata. Examples include exposed local paths (``\textit{/Users/md/Dropbox/dev/.../BTC-USD-15m.csv}'') or internal metadata (``\textit{betas=[output-128k-2025-02-19]}'').

(2) \textbf{Hard-coded Sensitive Parameter Smell}. Secrets or internal parameters that are embedded directly in rule files. For example, ``\textit{`budget\_tokens' :'120790'}''; hard-coded default values used as secrets, such as ``\textit{`SECRET\_KEY`: Used for signing JWT tokens. Defaults to `banana' if not provided}''. %, or ``\textit{api\_key = `my\_api\_key'}''.

(3) \textbf{Internal Resource Exposure Smell}. Internal or private resources that are exposed, revealing system structure or internal tooling. For instance, private workspace URLs such as ``\textit{https://www.notion.so/... backend ...450db1a?pvs=21}''.

(4) \textbf{Inconsistency Smell}. Contradictory or duplicated instructions. For example, one \cur\ instructs Cursor:``\textit{Use camelCase for component names (e.g., AuthWizard} [sic]\textit{)}'', %\olg{here example is like this or 'authWizard'?}\shu{Hi, Olga, yes, this is how the example shows. I just checked this cursorrules file again, and confirmed that it repeats `AuthWizard' twice}
and later: ``\textit{Use PascalCase for component names (e.g., AuthWizard)}''.

The presence of these smells has practical consequences.
Exposed paths can reveal project structure or private data, hard-coded parameters may encourage insecure handling of sensitive values, and internal resource links can leak information about internal documentation or system organization.
Conflicting instructions introduce uncertainty into the Cursor's behavior and increase the likelihood of unpredictable or inconsistent code generation, which is itself a security concern.

%\highlight{
\underline{\textbf{RQ4 Takeaway:}} Security-related guidance appears infrequently in \cur\ files, accounting for about 4.4\% of all coded instances. Potential security smells are also present but rare (1.5\%). %, including information exposure, hard-coded sensitive parameter, internal resource exposure, and inconsistency smell.
%}

\section{\uppercase{Discussion}}
\label{sec:discussion}

%\subsection{Discussion}
\textbf{Analysis of findings.} Our mixed-methods analysis shows that \cur\ files in open-source GitHub projects mostly occur in small, unpopular, single-maintainer projects that are not actively maintained. Many of these projects were likely used for testing the Cursor AI or educational purposes (i.e., \emph{toy} projects). A smaller fraction of the files occurs in popular and large repositories, indicating that these are active open-source projects strongly engaging with Cursor. Over time, \cur\ files have become slightly better maintained, with shorter update cycles and more localized modifications. Since this format was declared legacy in February 2025, we expect these files to become rarer, being replaced by more current prompt configuration formats.

Our qualitative content analysis shows that the themes and topics that organically emerged in \cur\ files also occur in the more recent \texttt{.mdc} files, and there is a topical continuity between the formats. Therefore, we believe that our qualitative analysis is also relevant for these more recent files. %\olg{to finish here}
The qualitative analysis results by \cite{context_engineering} and \cite{chatlatanagulchai2025agent}, which investigated \texttt{AGENTS.md} and other markdown-based files for AI agents, also reveal similar patterns: a larger focus on architecture and project structure with relatively lesser focus on issues like security.

\textbf{Threats to validity.}
\emph{Quantitative Research.}
%Our quantitative analysis and its external validity are limited by the nature of interacting with GitHub repositories.
%GitHub API does not retrieve complete results~\cite{akhoundali2025eradicating}, thus, the overall limitation of quantitative analysis is search completeness.
%Because our search strategy had to partition the query space to bypass GitHub's 1,000-result limit as suggested in~\cite{akhoundali2025eradicating}, some \cur\ files may not have been retrieved.
%In addition, the dataset is constructed from two snapshots collected in April and October 2025. While merging snapshots increases coverage, it also introduces uncertainty when interpreting temporal trends, as observed patterns may reflect snapshot timing rather than continuous evolution. Further, our analysis focuses exclusively on \cur\ files from public GitHub repositories, while private projects may have different patterns and behaviors.
Our quantitative analysis and its external validity are limited by the constraints of interacting with GitHub repositories.
Because the GitHub API does not return complete results~\cite{akhoundali2025eradicating} and our search strategy had to partition the query space to bypass the 1,000-result limit, some \cur\ files may not have been retrieved.
In addition, the dataset combines two snapshots (April and October 2025). Although this increases coverage, it also introduces some uncertainty when interpreting temporal trends.
Also, our analysis focuses exclusively on public repositories, while private projects may have different patterns and behaviors.

%\textbf{Construct Validity}.
%(1) Repository completeness and structural indicators.  
Regarding construct validity, we introduce a composite indicator of repository completeness and maturity. While its elements are commonly associated with well-documented and community-oriented projects, it should still only be interpreted as a coarse structural indicator.
%In addition, contributor counts rely on \texttt{mentionableUsers.totalCount} field of GraphQL, which may not perfectly match the intended construct of ``contributors''. 
Finally, repository-level attributes are capped, so activity in highly active repositories is only lower-bounded. % rather than precisely known.

\emph{Qualitative Research.}
%% 10.4.1 Techniques for Demonstrating Validity in Qualitative Studies.
%% Techniques for addressing threats to validity in qualitative research:
%%% # 1 Design Consideration: Developing a self-conscious research design. | Sampling decisions (i.e., sampling adequacy). | Employing triangulation. | Peer debriefing. | Performing a literature review. | Sharing perquisites of privilege.
%%%% # 1.1 Sampling decisions, i.e., sampling adequacy.
The key limitation of the qualitative study is our small sample size (65), which is not a representative sample of the whole dataset. 
This restricts the breadth of phenomena that can be observed and the representativeness of the resulting themes. While we observed coding stabilization on our sample, analyzing more files could potentially reveal new codes and themes.  %, and it also means that our findings may not generalize to private rule files that we cannot access.
Second, although the coding and theme construction followed the standards of thematic analysis, the resulting codebook and themes inevitably reflect the authors' interpretations.
%Different analysts might theme the codes differently.
These factors should be taken into account when interpreting the qualitative results. As is common in thematic analysis, its findings cannot be interpreted as generalizable to the whole population. 

\section{\uppercase{Related Work}}

%\textbf{Generative AI and prompt design in software engineering.}
LLMs are being applied to various software development tasks~\cite{dong2025survey,sergeyuk2026human}. %, as evidenced empirically by datasets such as 
%by the AIDev collection of open source projects where code has been written by AI agents
%AIDev~\cite{li2025rise} and %the PromptSet dataset that contains prompts shared in GitHub projects
%PromptSet~\cite{pister2024promptset} as well as self-reported by developers~\cite{otten2025prompting}. % Jamil et al.\cite{can_LLM_gen_HQcode_2025} provide evidence that LLMs with proper guidance can surpass human developers in generating high-quality code on certain tasks; however, AI-generated code does not yet perfectly pass all test cases~\cite{can_LLM_gen_HQcode_2025,promptE_or_fineTuning_2025}, and relying on AI-only feedback loops without human intervention can lead to quick degradation in security outcomes~\cite{shukla2025security}. Similarly, developers' perceptions of AI coding assistants are mixed: beginner practitioners may perceive LLMs as more helpful than more experienced programmers~\cite{lyu2025my,kudriavtseva2025my}. 
%We refer to the systematic literature reviews by \cite{dong2025survey} and \cite{sergeyuk2026human} for a more extensive analysis of the current literature on AI code generation and human-AI interaction in development environments, respectively.
Prompts can influence the quality of the produced code~\cite{della_prompt_code_quality,nam2025understandingsupportingdevelopersprompt}, and the recent literature actively investigates how to design effective prompts. \emph{Prompt engineering} refers to various prompt improvement techniques and methodologies that enhance LLM performance in the relevant areas and tasks~\cite{chen2025promptware,prompts_are_programms2,ronanki2025prompt,sahoo2024systematic}.%,schulhoff2025promptreportsystematicsurvey}. % For example, \cite{gao2023makes} show that sufficiently diverse demonstrations in the prompt improve performance in code intelligence tasks. % and provide guidance on how to select these samples. 
%At the same time, LLM fine-tuning to a specific task or knowledge remains a viable alternative to prompt engineering for improving the LLM performance~\cite{promptE_or_fineTuning_2025}.  %, while promptware engineering or prompt programming has recently emerged as a method inspired by established software engineering practices to produce and curate better prompts~\cite{chen2025promptware,prompts_are_programms2,prompt_with_me}. 

%\textbf{Prompt mining.}
Mining software repositories and analysis of open source projects is a very active research area~\cite{de2016systematic}. %Studies focus on very diverse aspects of software development and software project maintenance: for example, on policy-as-code adoption~\cite{Opdebeeck_2026PaC}, the CODEOWNERS feature usage~\cite{lulla2025automated}, attribution of contributions~\cite{holtgrave2025attributing}, code smells refactoring~\cite{shetty2026mapping}, commit activities~\cite{characterizing_commits_2022}, or the usefulness of a software composition analysis tool~\cite{nocera2025software}. 
Closest to our work, several studies have examined prompts in the context of software repositories and development practices. 

\emph{Prompt collections analysis.} \cite{siddiq2024fault} qualitatively examined prompts in 9 code generation benchmarks for writing quality issues. %They found that prompts frequently have poor quality, contain grammatical issues, and express the developer's intent ambiguously. 
\cite{understanding_prompt_management} analyzed 92 GitHub prompt repositories, quantitatively studying prompt sources, prompt storage practices, and prompt quality in terms of length, grammar, and readability. 

\emph{Prompts for specific development tasks.} \cite{promptingInWild} inspected the evolution of prompts from the PromptSet dataset~\cite{pister2024promptset}. \cite{promptingInWild} reports that prompt changes are usually limited to specific elements and are mostly introduced during feature development, and they are frequently not documented in the commit messages. \cite{rzig2025developer} report that PromptSet prompts frequently suffer from bias, vulnerability to injection attacks, and suboptimal performance. \cite{mao2025prompts} analyzed the PromptSet prompts and constructed a dataset of prompt templates from open source LLM apps. %, providing best practices for designing such templates. 
% Li et al.~\cite{prompt_with_me} worked with prompts curated in \cite{promptingInWild}.
Considering actual prompt usage by developers, \cite{desmond2024exploring} and \cite{nam2025understandingsupportingdevelopersprompt} qualitatively examined prompt editing behavior in an enterprise. %They found that users frequently did edits aimed at improving the prompt structure, and that some changes were introduced and then removed, potentially indicating the need for higher readability and the cognitive difficulties in remembering the effects of a prompt change. 
%Based on the analysis of developers' struggles with prompting, \cite{nam2025understandingsupportingdevelopersprompt} proposed AutoPrompt, a system to improve developers' prompts by augmenting them with inferred contextual information, including specifics, operationalization plan, localization/scope, code-base context, and developer's intent. 
\cite{zhong2025developer} characterized developers' interactions with ChatGPT, studying the most common tasks, programming languages, and the session structure, while \cite{siddiq2024quality} have shown that ChatGPT frequently produced low-quality code in these interactions.

\emph{Prompt configuration files analysis.} Closest to our study, 
% Jiang and Nam~\cite{jiang2025empirical} 
\cite{context_engineering} investigated the usage of AI configuration files in more mature open-source repositories, showing that only 5\% of repositories in their original sample contained such files. They identified 116 repositories containing \texttt{AGENTS.md} files and analyzed these files regarding the included headers, writing style, and the changes introduced to these files. Compared to our study, we examine a larger sample of \cur\   quantitatively, and we do not restrict ourselves to more established software projects, while our qualitative analysis did not start with any preconceived file structure as organized in the headers. \cite{chatlatanagulchai2025agent} further extend \cite{context_engineering} by collecting and analyzing \texttt{CLAUDE.md}, \texttt{AGENTS.md}, and \texttt{copilot-instructions.md} from projects shared in the AIDev dataset~\cite{li2025rise}. They found that these files evolve via frequent, small additions, and that, among the 16 available instruction types, developers prioritize functional context, while non-functional requirements like security and performance rarely receive attention. In our study, we observed that developers relatively frequently mention non-functional requirements such as code quality and maintainability, whereas security-related concerns receive comparatively little attention. The key novelty of our study is that we investigated \cur, a once-standard free-format prompt configuration files for Cursor. Our analysis reveals the emergence and usage of these files in open-source software projects. 

% to finish here by summarizing what we do differently
% In contrast to the aforementioned studies, we... 

%Mohsenimofidi et al.~\cite{context_engineering} 
%by extracting their section headings to explore what information developers provide and how they present it.
%Their results show that there is no established structure yet for content in \texttt{AGENTS.md} files, and that there is a lot of variation in how context is provided. % (descriptive, prescriptive, prohibitive, explanatory, conditional).
%Their results show substantial variation in content presentation, suggesting that an established structure has not yet been established.
%However, their analysis is limited to section headings and does not examine the textual content.
%To study how \texttt{AGENTS.md} files evolve, they manually review the commits associated with files that were modified more than 10 times. %including the source code diff, the commit messages, and any related issues or pull requests.
%They report no clear patterns in when and how often changes occur, but note that early modifications of AI configuration files tend to involve fine-tuning and adjusting instructions rather than major changes.
%Compared to our work, their analysis does not include fine-grained commit activities such as monthly commit trends.

\section{\uppercase{Conclusion}}

In this work, we have empirically investigated the \cur\ files in open-source GitHub repositories. In our mixed-methods study, we identified the behavior and properties of \cur\ files as part of open-source repositories, as well as the common themes and topics that developers mention in these prompt configuration files. Future research can build on our thematic analysis to perform a larger-scale study using LLMs to code all files in our dataset, as well as to extend the analysis to other prompt configuration file types. %\olg{Shuang, please mention in the repo (we need a Readme there) that the whole dataset will be shared once the paper is accepted.}\shu{Done.}

\bibliographystyle{apalike}
{\small
\bibliography{references}}

%\section*{\uppercase{Appendix}}
%If any, the appendix should appear directly after the references without numbering, and not on a new page. To do so please use the following command: \textit{$\backslash$section*\{APPENDIX\}}

\end{document}